# Spin-lattice relaxation in bismuth chalcogenides

*Robert E. Taylor[†], Belinda Leung[†], Michael P. Lake[†], and Louis-S. Bouchard[†‡§]*

[†]Department of Chemistry and Biochemistry, University of California, Los Angeles, Los Angeles, CA 90095, USA

[§]Department of Bioengineering and California NanoSystems Institute, UCLA, Los Angeles, CA 90095, USA

[‡] Corresponding author's email address:

bouchard@chem.ucla.edu

Last revised: July 15, 2012

## Abstract

Bismuth chalcogenides $Bi_2Se_3$ and $Bi_2Te_3$ are semiconductors which can be both thermoelectric materials (TE) and topological insulators (TI). Lattice defects arising from vacancies, impurities, or dopants in these materials are important in that they provide the charge carriers in TE applications and compromise the performance of these materials as TIs. We present the first solid-state nuclear magnetic resonance (NMR) study of the $^{77}$Se and $^{125}$Te NMR resonances in polycrystalline powders of $Bi_2Se_3$ and $Bi_2Te_3$, respectively. The spin-lattice ($T_1$) relaxation is modeled by at most two exponentials. Within the framework of this model, the NMR measurement is sensitive to the distribution of native defects within these materials. One component corresponds to a stoichiometric fraction, an insulator with a very long $T_1$, whereas the other component is attributed to a sample fraction with high defect content with a short $T_1$ resulting from interaction with the conduction carriers. The absence of a very long $T_1$ in the bismuth telluride suggests defects throughout the sample. For the bismuth selenide, defect regions segregate into domains. We also find a substantial difference in the short $T_1$ component for $^{125}$Te nuclei (76 ms) and $^{77}$Se (0.63 s) in spite of the fact that these materials have nearly identical lattice structures, chemical and physical properties. Investigations of the NMR shift and Korringa law indicate that the coupling to the conduction band electrons at the chalcogenide sites is much stronger in the telluride. The results are consistent with a stronger spin-orbit coupling (SOC) to the *p*-band electrons in the telluride. If most parameters of a given material are kept equal, this type of experiment could provide a useful probe of SOC in engineered TI materials.

**Introduction**

Bismuth chalcogenides have been of research interest for some time due to their thermoelectric (TE) properties. These semiconductor materials have narrow band-gaps, low thermal conductivity, and high thermoelectric figures of merit at room temperature.[1-4] Thermoelectric devices from such materials offer "long life, no moving parts, no emissions of toxic gases, low maintenance, and high reliability" but are limited by a "relatively low energy conversion efficiency".[5] Attempts to improve efficiency are aimed at fabricating nanostructures to reduce thermal conductivity by the interface and for quantum confinement of electrons.[6] Bismuth selenide ($Bi_2Se_3$) has been extensively studied and may be "considered as a model compound among thermoelectric materials for both experimental and theoretical analysis".[3] More recently bismuth chalcogenides have also been shown to be topological insulators (TI), which are novel quantum states of matter receiving much attention in the condensed matter physics community.[7-9] Topological insulators are characterized by insulating band gaps in the bulk while having a gapless surface state that is protected by time-reversal symmetry. The resultant properties may be useful in device applications.

Defects arising from lattice defects, impurities, or dopants in these materials are important in that they provide the charge carriers in thermoelectric applications and compromise the performance of these materials as topological insulators. For example, a common defect in $Bi_2Se_3$ is a selenium vacancy which contributes two electrons to the system.[10] Magnetic impurities such as Fe or Mn are known to open a gap at the Dirac point in topological insulators.[9] A better understanding of the defects leading to suboptimal insulating properties requires methods to interrogate mechanisms that lead to conduction. Nuclear magnetic resonance (NMR) techniques can probe coupling of specific lattice site nuclei to the conduction band electrons in semiconductors and metals. Such site-specific measurements could be used to probe the involvement of local orbitals in the conduction process.

The purpose of this study is to provide a NMR characterization of the polycrystalline powders used as starting materials for growing single crystals, thin films, and nano-powder preparations of $Bi_2Se_3$ and $Bi_2Te_3$ and for preparing selectively doped versions of these compounds. The characterization of these polycrystalline starting materials provides the background for an understanding for the effect of native defects on the NMR spin responses. The spin-lattice relaxation times ($T_1$) as well as the observed shifts of the $^{77}$Se nuclei and $^{125}$Te nuclei can be measured as function of temperature. Such information is useful when more traditional measurements, e.g., transport, are not readily available from powdered samples.[10] This information should prove useful in future investigations of single crystals and thin films, in the comparison of the effects of native defects with those of dopants, and in the investigation of surface effects on the NMR responses as well as defects in nanoscale preparations of these materials.[11]

**Experimental Methods**

The $^{77}$Se NMR data were acquired at ambient temperature with a Bruker DSX-300 spectrometer operating at a frequency of 57.30 MHz. The $^{125}$Te resonance frequency was 94.79 MHz. NMR data on static polycrystalline samples were acquired using a standard Bruker X-nucleus wideline probe with a 5-mm solenoid coil. Samples of $Bi_2Se_3$ and $Bi_2Te_3$ were purchased from Alfa Aesar and used without further purification or recrystallization. The polycrystalline $Bi_2Se_3$ and $Bi_2Te_3$ samples were ground with 50 weight % NaCl to prevent particle contact of the grains and thus avoid radiofrequency (RF) skin-depth issues. The sample was confined to the length of the RF coil. The $^{77}$Se and $^{125}$Te π/2 pulse widths in the wideline probe were 4 μs.

Spectral data were acquired using a spin-echo sequence [π/2)$_x$ – τ – π)$_y$ - *acquire*]. The echo delay, τ, was set to 10 μs. The spin-echo sequence is useful in minimizing pulse ringdown effects and for the recovery of the chemical shielding interaction in the presence of heteronuclear dipolar coupling. Magic-angle spinning (MAS)[12-14] spectra were acquired on the same samples with a standard Bruker MAS probe using a 4-mm outside diameter zirconia rotor with sample spinning rates of 10 kHz.. The $^{77}$Se and $^{125}$Te π/2 pulse widths for the MAS experiments were 4.4 μs. Simulations of the shielding powder patterns for the spectra obtained from the static samples were performed with the solids simulation package (''solaguide") in the TopSpin (Version 3.0) NMR software program from Bruker BioSpin.

Data for determining the spin-lattice relaxation times ($T_1$) were acquired with the saturation-recovery technique.[15] The $^{77}$Se and $^{125}$Te chemical shift scales were calibrated using the unified Ξ scale[16], relating the nuclear shift to the $^1$H resonance of dilute tetramethylsilane in $CDCl_3$ at a frequency of 300.13 MHz.

**Results and Discussion**

Bi$_2$Se$_3$ and Bi$_2$Te$_3$ crystallize in the rhombohedral space group $D_{3d}^5$ ($R\overline{3}m$) with three formula units per unit cell.[7] This is an ABC-type structure consisting of five atomic layers forming a quintuple layer with the order Se-Bi-Se-Bi-Se or Te-Bi-Te-Bi-Te in the respective materials. The coupling is strong between two atomic layers within the quintuple layer, with the quintuple layers bound by weaker van der Waals interactions.

The majority of elemental and binary semiconductors typically crystallize in one of four highly symmetrical crystal structures[17]: diamond, zinc blende, wurtzite, or rock salt (NaCl). The NMR spectra typically reflect the high symmetry found in these materials. However, as this high symmetry is not present in the $D_{3d}^5$ ($R\overline{3}m$) space group for the Bi$_2$Se$_3$ and Bi$_2$Te$_3$ structures, the NMR spectra are more complicated. For example, the lack of cubic symmetry may give rise to anisotropy in the chemical shift[18,19] and Knight shift.[20] In addition, there are two crystallographically distinct sites for Se and Te in the quintuple layers of Bi$_2$Se$_3$ and Bi$_2$Te$_3$ lattices.

The $^{77}$Se and $^{125}$Te wideline NMR spectra of static samples of Bi$_2$Se$_3$ and Bi$_2$Te$_3$ acquired with a spin-echo technique are shown in Figure 1. Analyses of these spectral lineshapes prove difficult due to the number of nuclear interactions present. The observed resonance arises from both the chemical shift and the Knight shift in these conducting samples. Both the chemical shift and Knight shift can have anisotropic contributions due to the lower symmetry of the crystal structure. As previously noted, there are also two distinct crystallographic sites for the Se and Te nuclei in these compounds. A further complication due to defects may also produce asymmetrical resonances. As an example, asymmetrical resonances for spin-½ $^{207}$Pb have been observed in the semiconductor PbTe, which has the rock salt (NaCl) structure with cubic symmetry.[21] This asymmetry arises from an inhomogeneous distribution of defects with the increased charge carrier concentrations having different Knight shifts. Subsequent work[22] has shown that the $^{207}$Pb Knight shift in PbTe scales as $n^{1/3}$, where $n$ is the charge carrier concentration.

Magic-angle spinning (MAS) at 10 kHz in the magnetic field strength used in this study does not narrow the spectral resonances at all in comparison to the wideline spectra of the static samples for either sample. The lack of resonance narrowing by MAS of a spin-½ nucleus next to a quadrupolar nucleus has been reported[23] for $^{199}$Hg in the semiconductors $HgBr_2$ and $HgI_2$ at this magnetic field strength. Faster MAS at 22 kHz in a higher magnetic field has yielded only a marginal narrowing of the $^{125}$Te resonance in $Ag_{0.53}Pb_{18}Sb_{1.2}Te_{20}$ over that observed in the spectrum of the static sample.[24] The fact that the resonances do not break into sidebands upon MAS indicates that the spectral broadening is not inhomogeneous. The resonance lineshape cannot be due to a simple distribution of chemical shifts (or Knight shifts) resulting from an orientational dependence. The various contributions to the spectral resonances cannot be separated with MAS at this magnetic field strength.

A previous study[25] of conducting transition-metal ditellurides by $^{125}$Te NMR used a single tensorial interaction to simulate the lineshapes rather than trying to separate the chemical and Knight shift contributions. That same approach has been used here. The solid lines in Figure 1 are the results from simulations to determine the principal components of the shielding tensor. The extracted parameters of this interaction are given in Table I. The simulations do not produce ideal fits to the experimental spectra due to the presence of the other interactions, such as the inhomogeneous distribution of defects, in these materials. For that reason, no error analysis is provided. Nevertheless, the simulations provide a quantitative framework for the description and comparison of the $^{77}$Se and $^{125}$Te resonances in these materials.

The chemical shift powder pattern shown in Figure 1 for $Bi_2Te_3$ extends over a range of approximately 1,000 ppm. To show that the full spectral resonance could be acquired in a single spectrum without rolloff due to the RF excitation, the $^{125}$Te *v*ariable *o*ffset *c*umulative *s*pectra (VOCS)[26] are shown in Figure 2a. The summation of the cumulative spectra is overlayed with the single $^{125}$Te spectrum of the $Bi_2Te_3$ to illustrate that the full spectrum can be uniformly excited with the acquisition conditions used in this investigation. In a similar way, a single RF excitation was sufficient to acquire the full spectral resonance of $^{77}$Se nuclei in $Bi_2Se_3$.

The chemical shift powder pattern was independent of temperature over the range 150-400 K. Knight shifts are not strongly dependent upon temperature for semiconductors having sufficient carrier concentrations so that the semiconductors are in a metallic regime having degenerate and independent charge carriers.[17] The shielding anisotropy in bismuth selenide is much larger than for bismuth telluride. The spin-lattice relaxation results (*vide infra*) suggest that this higher anisotropy likely arises from a more inhomogeneous distribution of defects within the bismuth selenide. Also, the isotropic shift for bismuth telluride is about tenfold larger than for bismuth selenide. This larger value in the isotropic shift for the bismuth telluride is also reflected in the differing spin-lattice relaxation times (*vide infra*).

The spin-lattice ($T_1$) relaxation data for $^{125}$Te nuclei in $Bi_2Te_3$ can be fit with a single exponential, yielding a $T_1$ of 76 ms (Figure 3). For $^{77}$Se nuclei in $Bi_2Se_3$, the spin-lattice relaxation clearly could not be adequately described with a single exponential fit. A biexponential fit (Figure 4) yielded two distinct $T_1$ components of roughly equal magnitude: one with a time constant of 0.63 s and the second with a $T_1$ greater than 60 s. A stretched exponential did not provide as good a fit to the data as the two component model. We verified that the biexponential recovery of the saturation recovery data for the $^{77}$Se was not due to relaxation anisotropy across the spectral resonance. This is illustrated in Figure 5 which shows spectra from the saturation recovery experiment acquired with delays of 1, 5, and 120 s after saturation of the resonance.

The simplest model for the existence of two relaxation components in the bismuth selenide is an inhomogeneous distribution of native defects. These samples are prepared by a solid-state reaction of bismuth metal with either selenium or tellurium metal in a high temperature furnace.[27] Powder X-ray diffraction (PXRD) shows that both the bismuth selenide and bismuth telluride are crystalline (Figure 6). However, our results suggest that not only NMR can probe specific sites but it is sensitive to the distribution of native defects within these materials. The NMR technique thus reveals complementary information to the PXRD spectrum. Within the framework of this model, one component would be the existence of a stoichiometric $Bi_2Se_3$ fraction, an insulator with a very long $T_1$, whereas the other component could correspond to a sample fraction with high defect content with a short $T_1$ resulting from interaction with the conduction carriers. The absence of a very long $T_1$ in the bismuth telluride suggests defects throughout the sample.

Transport measurements are not readily made on powdered samples. However, according to prior studies[10,28], the carrier concentrations for both the bismuth telluride and selenide are typically around $2\times10^{25}$ m$^{-3}$. The difference is that defects in bismuth telluride tend to produce holes while those in bismuth selenide tend to produce electrons. The accepted model[28] for defects in the telluride is that the excess component substitutes randomly in the lattice sites of the other, an "anti-structure" or "anti-site" defect. Bismuth in a Te site creates a hole while Te in a Bi site creates an electron. For $Bi_2Se_3$, the common defect is Se vacancies[10], creating electrons. While the Se vacancy model has successfully accounted for most of the observed experimental results, the co-existence of some anti-site defects has been more recently proposed.[29] The charge carrier concentration arising from these defects in these two compounds is usually quite similar.[10,28] Knight shifts scaling with the carrier concentration would then be expected to be similar in both compounds. For example, the nuclear relaxation rates, $1/T_1$, for both $^{207}$Pb and $^{125}$Te have been shown to be directly proportional to the charge carrier concentration for both PbTe and $Ag_{0.53}Pb_{18}Sb_{1.2}Te_{20}$.[24]

The fact that $Bi_2Te_3$ contains only a single relaxation component whereas $Bi_2Se_3$ has two suggests that anti-site defects dominate in $Bi_2Te_3$ whereas in $Bi_2Se_3$ the defect regions segregate into domains. The relaxation times for defect-heavy fractions in both samples can be compared. The short $T_1$s of $Bi_2Te_3$ (76 ms) and of $Bi_2Se_3$ (0.63 s) suggest the presence of a strong coupling to the conduction band electrons in these materials. This difference in relaxation times by an order of magnitude in these two materials is surprising at first sight because two materials have nearly identical lattice structures, chemical and physical properties, i.e., both are thermoelectrics and topological insulators under appropriate conditions. In addition, the NMR-active nuclei[16] in both samples are similar in terms of magnetic properties and natural abundance. The quadrupolar $^{209}$Bi with a spin of 9/2 has a natural abundance of 100%. The only NMR-active isotope of selenium is the spin-½ $^{77}$Se with a natural abundance of 7.63%. Tellurium has two spin-½ isotopes: $^{125}$Te with a natural abundance of 7.07% while $^{123}$Te has a much smaller natural abundance of 0.89%. As noted earlier, the nuclear relaxation rate is directly proportional to the charge carrier concentration. However, as also noted earlier, the charge carrier concentrations in these powdered materials are expected to be similar. Despite these similarities, the relaxation times for the defect-heavy fractions in both samples differ by about an order of magnitude.

A possible explanation of the difference in relaxation times is as follows. The expression below is for hyperfine coupling,[30]

$$\mathcal{H}_{HF} = 2\gamma\hbar\mu_B \boldsymbol{I} \cdot \left[\frac{\boldsymbol{l}}{r^3} - \frac{\boldsymbol{s}}{r^3} + \frac{3\boldsymbol{r}(\boldsymbol{s}\cdot\boldsymbol{r})}{r^5} + \frac{8}{3}\pi\boldsymbol{s}\delta(\boldsymbol{r})\right]$$

where $\gamma$ is the gyromagnetic ratio, $\mu_B$ is the Bohr magneton, $\boldsymbol{I}$ is the nuclear spin, $\boldsymbol{l}$ is the orbital moment of the electron, $\boldsymbol{s}$ is the electronic spin, and $\boldsymbol{r}$ is the distance from the nucleus. The last term is the contact interaction, which contributes nothing if the unpaired electronic wavefunction possesses no *s*-band character. The two middle terms are the electron-nuclear dipole-dipole interaction. The first term is the interaction of the nuclear moment with the electron's orbital moment. Two possible reasons for the coupling to the conduction band elections would be the presence of *s*-band character or a substantial contribution from spin-orbit coupling. While electronic structure calculations for the conduction band electrons do not show significant differences between these two materials (see Figure 4 of Ref. [31]), they do show that *s*-band electrons are only a negligible fraction of the electron density of states across the conduction band. This leaves the contribution from spin-orbit coupling.

Further evidence of coupling to the conduction band electrons was obtained from $T_1$ measurements over the temperature range 173-423 K. Korringa-law behavior[32] was observed in the low temperature range for $Bi_2Te_3$, whereas above 320 K, the relaxation data indicate thermally-activated relaxation arising from interaction with the charge carriers (e.g. possibly interband excitations and not to be confused with the band gap between the valence and conduction bands). The activation energy for this high temperature process is 8.44 kJ/mol (87 meV). The results for $Bi_2Te_3$ are shown in Figure 7. This "unusual temperature dependence of the spin-lattice relaxation" in which metallic behavior as indicated by the Korringa-law behavior observed at lower temperatures with the "activationlike temperature dependence […] characteristic of semiconductors" at higher temperatures has been previously observed in $Tl_2Se$.[33] The $^{125}Te$ $T_1$ results presented here were treated in this same manner.

For good conductors, the nuclear spin-lattice relaxation is usually dominated by interaction with the conduction band electrons. For this interaction, Korringa[32] has shown that the product of the spin-lattice relaxation time with the temperature ($T_1T$) is a constant. Given the inhomogeneous distribution of defects within $Bi_2Se_3$ giving rise to a multicomponent spin-lattice relaxation recovery and the exceedingly long spin-lattice relaxation times for this compound, similar experiments were not run on the powdered $Bi_2Se_3$. Such experiments will be pursued in future work on single crystals of this material. We have, however, carried out measurements of $T_1T$ versus temperature on Fe-doped $Bi_2Se_3$ and found no evidence of a Korringa law.[33]

As mentioned earlier, an explanation for the large difference in the $^{125}Te$ and $^{77}Se$ relaxation times could be spin-orbit coupling (SOC). An estimate based on isolated Te and Se atoms yields a SOC constant for the valence electrons that is 2.8 times larger for Te. However, the spin-lattice relaxation rate is proportional to the square of the interaction matrix element (in this case, the SOC), yielding a factor of 7.8 which scales the $T_1$ of 76 ms to 0.6 s, reasonably close to observed value of 0.63 s. A large SOC can lead to an appreciable hyperfine coupling to the conduction band electrons. Due to the heavy constituent elements in $Bi_2Te_3$, SOC effects have been shown to be very important.[34] Zhang, *et. al.*[7] have pointed out the role that SOC plays in topological insulators, noting that "the guiding principle is to search for insulators where the conduction and the valence bands have the opposite parity and a 'band inversion' occurs when the strength of some parameter, say the SOC is tuned". Detailed studies involving the production of doped and variable defect-content samples are underway to determine the origin of these relaxation components.

**Conclusions**

The $^{77}$Se and $^{125}$Te resonances and spin-lattice relaxation times of polycrystalline powders of $Bi_2Se_3$ and $Bi_2Te_3$ have been measured. The lower symmetry in the structures of these materials presents significant challenges in unraveling all the interactions constituting the resonance lineshapes. Below ambient temperature, a Korringa relationship indicating interaction with the charge carriers was found in $Bi_2Te_3$. Above ambient temperature, an activation energy of 8.44 kJ/mol (87 meV) was determined. An inhomogeneous distribution of defects gave rise to a multicomponent spin-lattice relaxation recovery for $Bi_2Se_3$. A good understanding of the role of defects and impurities in these materials is required to elucidate their intrinsic physical properties and improve their performance as TEs or TIs. And to this end, NMR is expected to be an important probe of site-specific conductivity and electronic structure. In particular, spin-lattice relaxation could be a useful probe of SOC in engineered TI materials.

**Acknowledgments**

Financial support from DARPA is acknowledged. L.-S. B. thanks M.G. Kanatzidis, X. Kou, J. Shih, K.L. Wang, S.D. Mahanti, F. Xiu, Y. Fan and L. He for useful discussions.

**Tables**

**Table I.**

| $^{77}$Se and $^{125}$Te Principal Tensor Components of the NMR shift of Bi$_2$Se$_3$, and Bi$_2$Te$_3$ | | | | | | | | |
|---|---|---|---|---|---|---|---|---|
| Static Sample[e] | $\delta_{11}$ | $\delta_{22}$ | $\delta_{33}$ | $\delta_{iso}$ | $\zeta_{csa}$[a] | $\eta$[b] | $\Omega$[c] | $\kappa$[d] |
| Bi$_2$Se$_3$ | 379 | 204 | -466 | 38.8 | -505 | 0.348 | 846 | 0.58 |
| Bi$_2$Te$_3$ | 724 | 461 | -56 | 376 | -432 | 0.607 | 780 | 0.33 |
| [a] $\zeta_{csa} = \delta_{33} - \delta_{iso}$. <br> [b] $\eta = (\delta_{22} - \delta_{11})/\zeta_{csa}$. <br> [c] $\Omega = \|\delta_{33} - \delta_{11}\|$. <br> [d] $\kappa = 3(\delta_{22} - \delta_{iso})/\Omega$. <br> [e] Shifts are in part per million (ppm) and referenced using the $\Xi$ scale, relating the $^{77}$Se and $^{125}$Te shift to the $^1$H resonance of dilute tetramethylsilane in CDCl$_3$ at a frequency of 300.13 MHz. (Ref. 16). | | | | | | | | |

**Figures and Figure Captions**

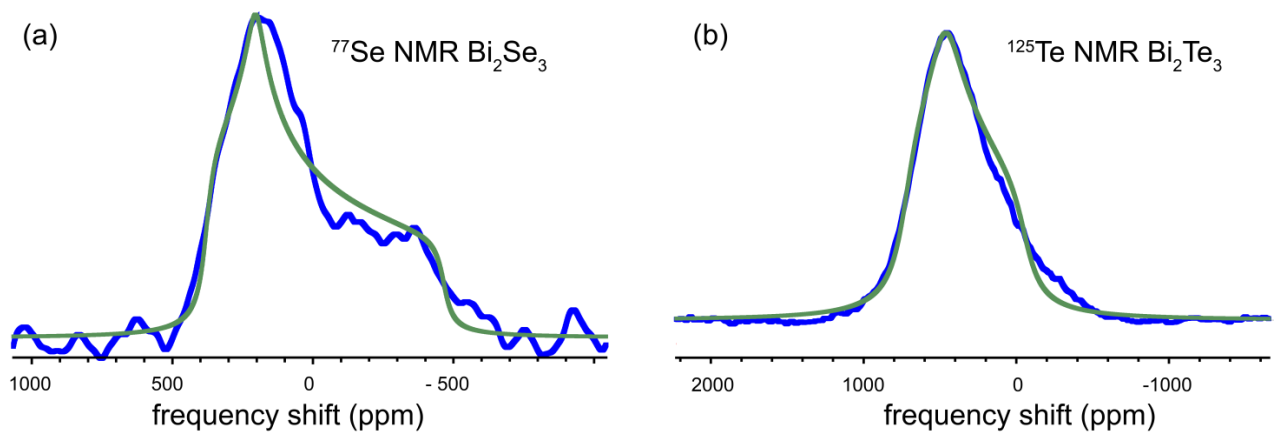

**Figure 1.** $^{77}$Se and $^{125}$Te powder spectra for (a) Bi$_2$Se$_3$ and (b) Bi$_2$Te$_3$ indicate much higher anisotropy in the bismuth selenide. The mean frequency shift for bismuth telluride is about tenfold larger than for bismuth selenide. The fitted values of the NMR shift tensor are listed in Table I.

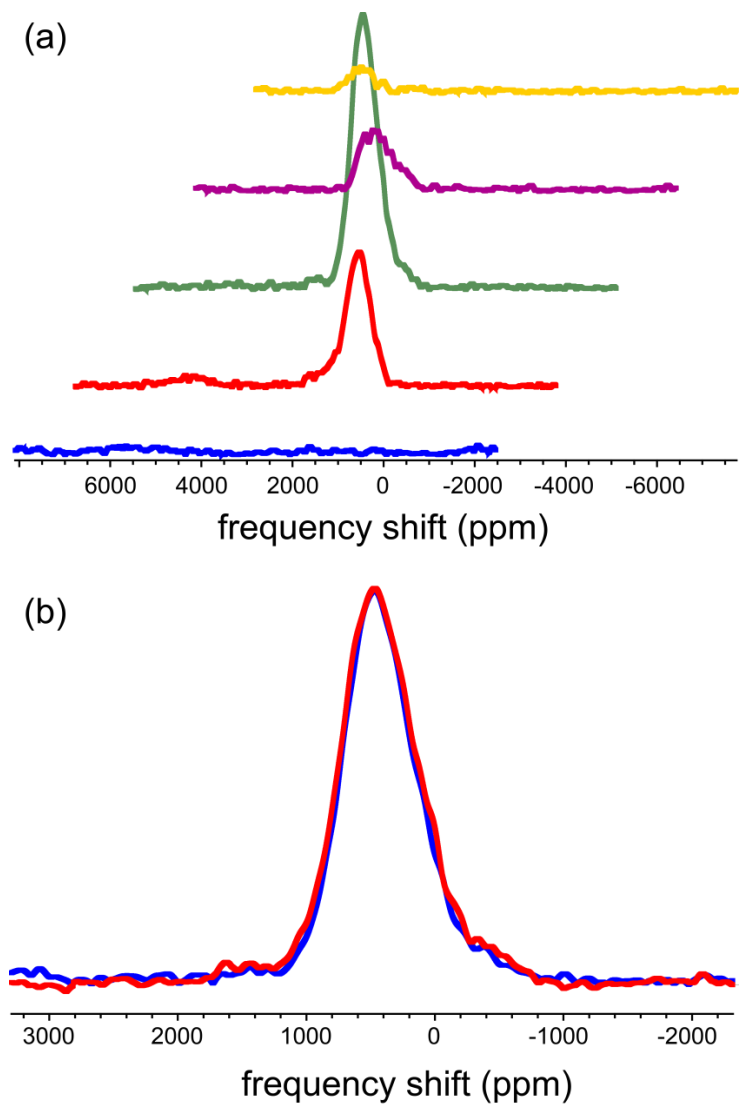

**Figure 2.** (a) $^{125}$Te VOCS subspectra acquired at different RF transmitter offsets and (b) overlay of VOCS summation (blue trace) with the single spectrum of $Bi_2Te_3$ to (red trace) illustrate that the $^{125}$Te spectrum can be uniformly excited and acquired as a single spectrum.

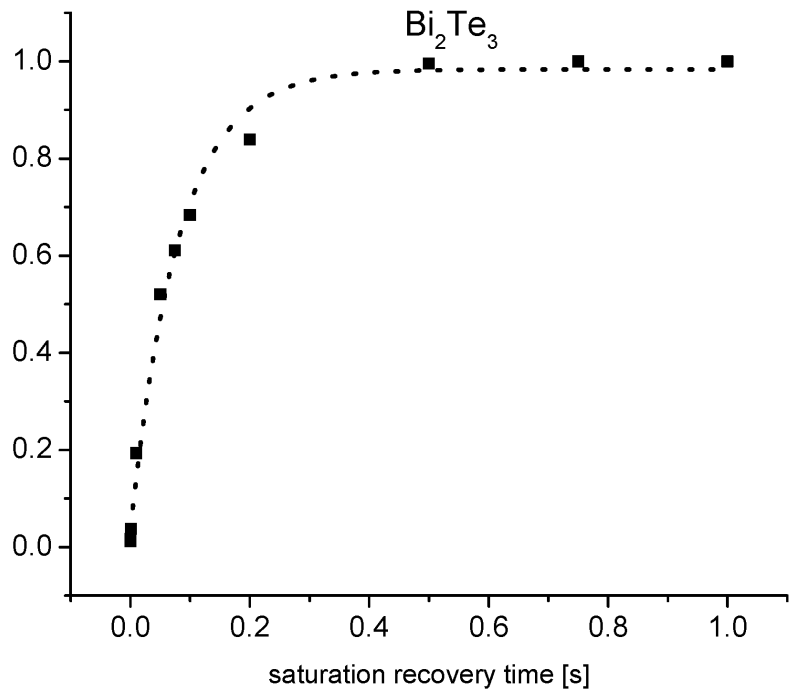

**Figure 3**. The spin-lattice saturation recovery ($T_1$) relaxation data for $^{125}$Te nuclei in Bi$_2$Te$_3$ at 297 K. The dotted line shows a fit of a single exponential with a time constant of 76 ms.

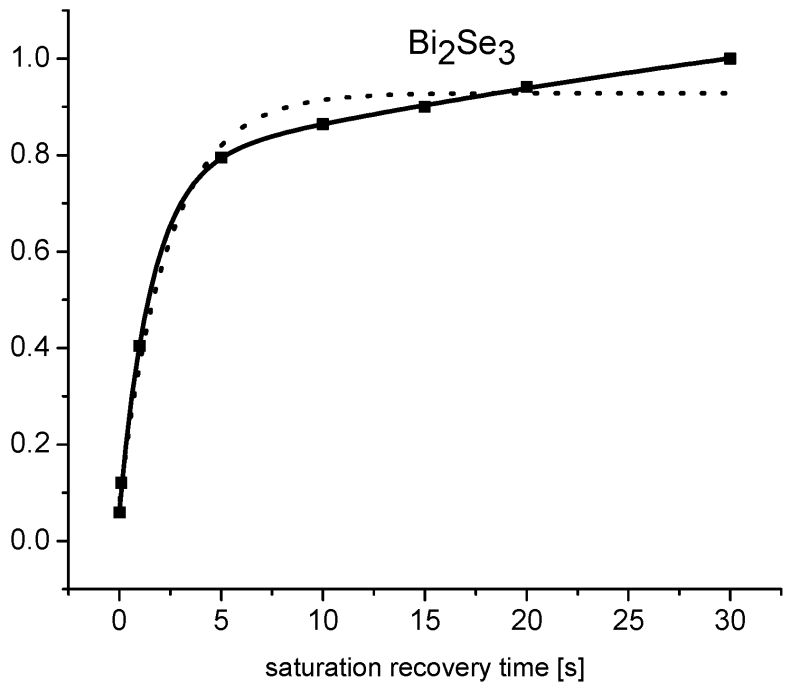

**Figure 4**. The spin-lattice saturation recovery ($T_1$) relaxation data for $^{77}$Se nuclei in Bi$_2$Se$_3$ at 297 K. The solid line shows a fit of a biexponential with a time constants of 0.63 s and greater than 60 s. The dotted line is a fit to a single exponential, outlining the need for two components.

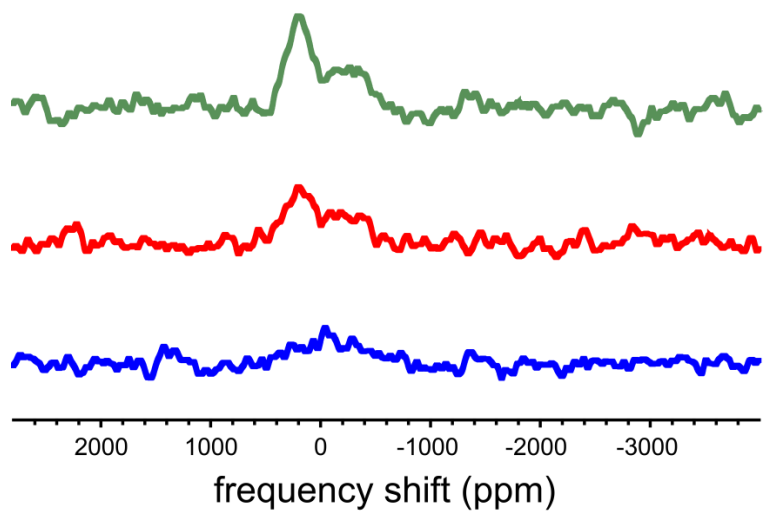

**Figure 5**. Spectra from the $^{77}$Se spin-lattice saturation recovery ($T_1$) relaxation experiment for Bi$_2$Se$_3$ at 297 K with delays of 1s (bottom), 5 s (middle), and 120 s (top) after the saturation sequence.

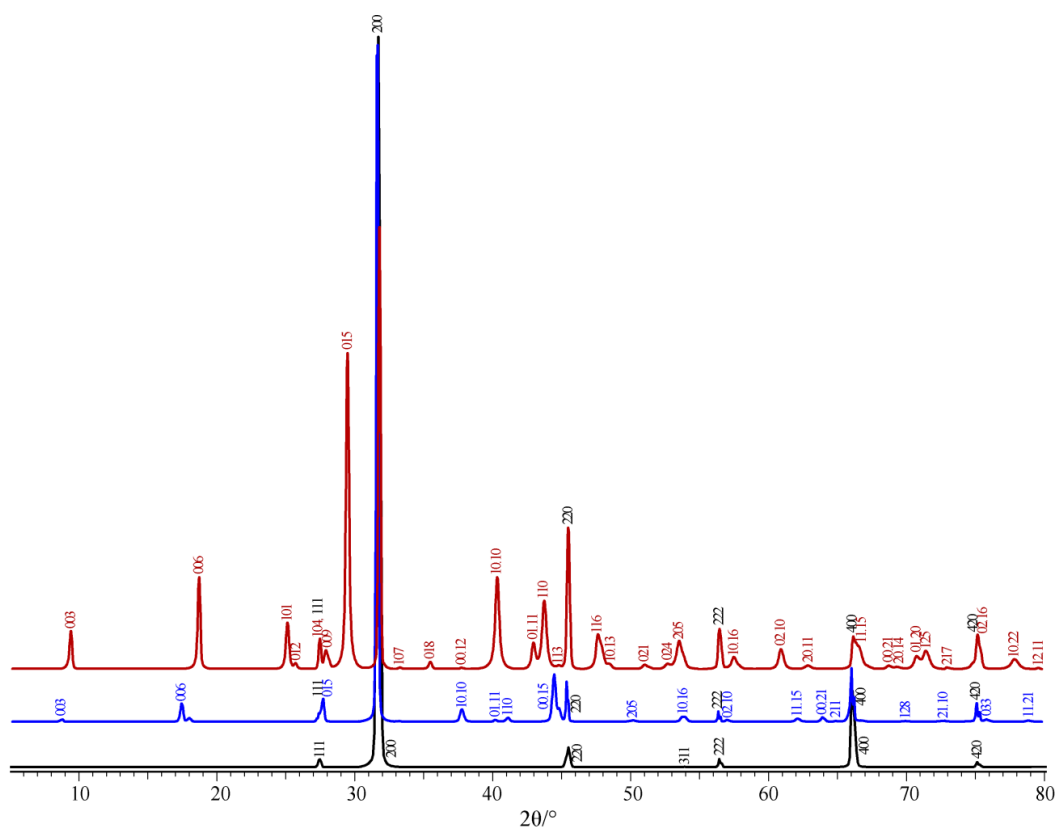

**Figure 6**. Powder X-ray diffraction (PXRD) spectra of bismuth selenide and bismuth telluride. The upper trace (red) is for $Bi_2Se_3$. The middle trace (blue) is for $Bi_2Te_3$. Both samples were ground with NaCl and therefore contain some amounts of NaCl. The PXRD spectrum for NaCl is shown in the lower trace (black).

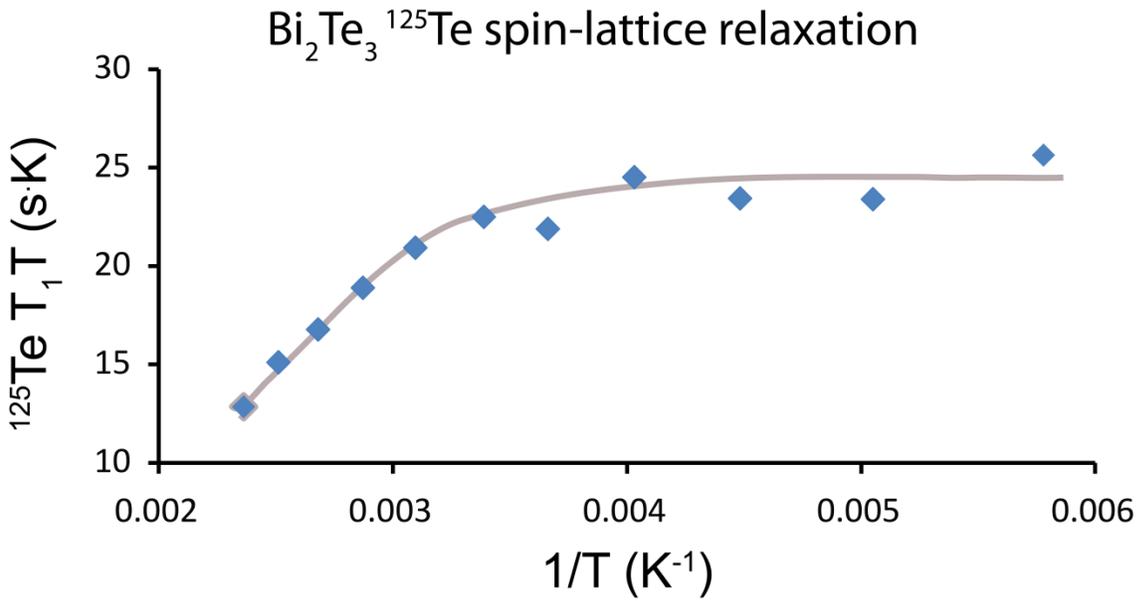

**Figure 7.** $T_1$ of $^{125}$Te as function of temperature for Bi$_2$Te$_3$ shows thermal activation at high temperature and Korringa law behavior at low temperatures, where the product $T_1 \cdot T$ becomes constant.

**Table Of Content (TOC) Graphic**

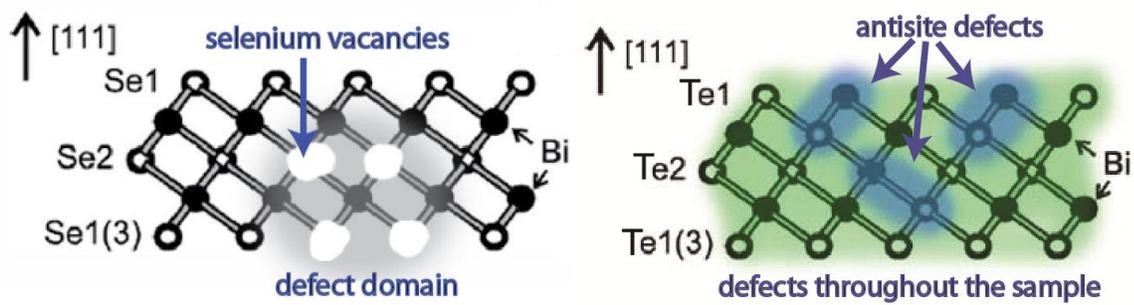